\documentclass[preprint]{aastex}
\usepackage{epsfig}
\shorttitle{Companions to HD 141569}
\shortauthors{Weinberger et al.}

\begin{document}

\renewcommand{\topfraction}{0.9}
\renewcommand{\bottomfraction}{0.9}
\renewcommand{\floatpagefraction}{0.9}

\title{Stellar Companions and the Age of HD 141569 and its
Circumstellar Disk}

\author{A. J. Weinberger\altaffilmark{1}, R. M. Rich\altaffilmark{1},
E. E. Becklin\altaffilmark{1}, B. Zuckerman\altaffilmark{1},  and
K. Matthews\altaffilmark{2}}

\email{alycia@astro.ucla.edu}

\altaffiltext{1}{Department of Physics and Astronomy, University of
California Los Angeles, Box 156205, Los Angeles, CA 90095;
alycia,rmr,becklin,ben@astro.ucla.edu}

\altaffiltext{2}{Palomar Observatory, California Institute of
Technology, 320-47, Pasadena, CA 91125;
kym@caltech.edu}

\begin{abstract}
We investigate the stellar environment of the $\beta$ Pictoris-like star
HD 141569 with optical spectroscopy and near infrared imaging.  The B9.5
Ve primary, and an M2 and an M4 type star both located in projection
$<$9$''$ away have the same radial velocity and proper motion and
therefore almost certainly form a triple star system.  From their x-ray
flux, lithium absorption and location on pre-main-sequence evolutionary
tracks, the companions appear to be 5 Myr old.  HD 141569 A is now one
of the few stars with a circumstellar disk that has a well-determined
age. The circumstellar disk is composed of secondary debris material
(Fisher et al. 2000), thus placing an upper limit on the time needed for
disk processing.  These three stars may be part of an association of
young stars.

\end{abstract}


\section{Introduction}

Stars with far-infrared excesses generated in circumstellar disks were
discovered in abundance by the IRAS mission. Since circumstellar
material disappears over time due to processes such as stellar winds,
radiation pressure, and accretion onto stars and planetesimals, most
infrared excess stars are young, $\lesssim$10$^8$~yr
\citep{Habing99,Spangler99,SilverstonePhD}.  To develop an understanding of the
evolution of circumstellar material and its relationship to the
formation of planets, detailed studies of individual disks must be
combined with information about stellar ages.  However, 
stars which are nearing the main sequence are difficult to age
because they lack traditional indicators such as Lithium and
x-ray emission and they move through the color-magnitude diagram very
quickly.
  
The Herbig AeBe stars (HAEBEs) are thought to be in this transition
phase, intermediate in evolution between protostars and stars on the
zero age main sequence (ZAMS).  The classical definition of HAEBEs,
i.e. that they are of spectral type A or earlier, are found in clouds,
and show emission lines and reflection nebulae \citep{Herbig} almost
certainly selects for such young objects.  In addition, more recent
studies \citep{The94} use the presence of near or far-infrared excess
to select HAEBEs.  Thus, objects such as HD 141569, a B9.5~Ve star with
H$\alpha$ in emission and 12--100 \micron\ excess, which are not
associated with any cloud or reflection nebula, fall into the HAEBE
class.  There have been attempts to date HAEBE stars within this general
picture based on the strength of their infrared excesses
\citep{Hillenbrand92}, and authors have proposed an evolutionary sequence
from the embedded HAEBEs to the isolated $\beta$ Pictoris or Vega-like
main sequence stars \citep[e.g.]{malfait}.

The presence of lower mass companions to these objects represents an
independent way of measuring their age, assuming they and the
companions are coeval.  Lower mass companions may not yet be on the ZAMS
and their ages can be estimated from theoretical pre-main sequence
evolutionary tracks.  For early-type stars near the ZAMS which also have
circumstellar dust, this is probably the most accurate way to determine
the disk age.  The well-studied disk stars, $\beta$ Pic, Fomalhaut, and
HR 4796A, were all dated by their association with lower mass co-moving
companions \citep{barrado,barrado97,stauffer}.

HD 141569 is joining the ranks of well-studied dusty disk stars with the
discovery of its large and morphologically complicated debris disk
\citep{Silverstone00,Weinberger99,Augereau99,Fisher00}.  In 1938,
Rossiter identified HD 141569 as a member of a potential triple system
with a second star 7.$''$5 away (B) and a third star 1.$''$5 from it (C)
\citep{Rossiter43}.  In a study of HAEBEs, \citet{Gahm83} measured the
spectral type of HD141569B as G0~V, and \citet{Lindroos85} concluded,
based on its magnitude and color, that it was a background star.
However, \citet{Gahm83} noted the presence of ``peculiar'' emission
lines in the spectrum of HD 141569B, including Ca H and K and H$\beta$,
which are often associated with young stars.

The parallactic distance to HD 141569A was measured by the Hipparcos
mission as 99$\pm$10 pc, which combined with its visual magnitude of
6.8, makes it underluminous for its spectral type, just as are other
young A-type infrared excess stars such as $\beta$ Pic, 49 Ceti and HR
4796A \citep{jura,lowrance00} (see also Figure \ref{fig_HR}).

Given the apparent proximity of two stars to HD 141569A, and given that
the disk around HD 141569A extends to half of the projected distance
between HD 141569A and B, this system could be dynamically interesting
if the three stars are physically associated.  In this paper, we show,
via comparison to previous astrometry plus new near infrared imaging and
visual spectroscopy, that HD 141569 A/B/C form a triple system, and we
estimate their ages.

\section{Observations and Data Analysis}

\subsection{HST and Ground-based Imaging}
On 1998 September 27, Short integration time images of the HD 141569
system were obtained with the NICMOS camera 2 on the Hubble Space
Telescope in order to acquire the primary star for coronagraphic
observations. Two simple ACCUM mode images of 0.342 s each were taken
with the F171M filter ($\lambda_{eff}$=1.721, FWHM=0.071) and
ten dark frames of the same integration time were taken two days earlier
for calibration of those acquisition images.  The two source images were
used for cosmic ray rejection by taking the minimum of the two values
for every pixel.  The median of the dark images was subtracted from the
science image to remove the effects of detector shading and bias.  The
resulting image was divided by an on-orbit flat field frame taken in the
same filter.  The best available photometric calibration was applied in
which 0~mag=948 Jy and 1 adu s$^{-1}$ = 1.071$\times$10$^{-5}$ Jy.

Near-infrared J ($\lambda_c$=1.27, $\Delta\lambda$=0.25), H
($\lambda_c$=1.65, $\Delta\lambda$=0.32), K ($\lambda_c$=2.20,
$\Delta\lambda$=0.40), and L$_s$-band ($\lambda$=3.45,
$\Delta\lambda$=0.57) images were taken with the Hale 200-inch
Telescope on 1999 May 25 and 28.  The infrared camera had a pixel scale
of 0.$''$125 per pixel and a full field of view of 32$'' \times$ 32$''$.
The night of 25 May was cloudy, but the flux ratios of B and C to star A
were measured at J, H, and K bands.  Short integration time, 0.374 s,
images were taken in which all three stars appear in every full-field
frame.  The seeing was 0.$''$82 at K band, so stars B and C were easily
resolved.  So as not to saturate HD 141569A at J and H bands, the
chopping secondary was used to smear the light in a direction
perpendicular to the position angle (PA), 301.9\degr\ between components
B and C.

The night of 28 May was photometric and measurements of
HD 141569A at all four wavelengths were preceded by measurements of the
photometric standard HD 129655.  For both target and standard, an 8$''$ square
subframe of the full array was employed to allow fast readout
and prevent the images from saturating the bright primary at J and H
or the thermal background at L$_s$.  At each filter, 100 integrations of
0.07~s were coadded for a total on-source integration time of 7~s.  One
0.54~s image of HD141569 B and C was also obtained at L$_s$.

Sky frames of the same integration time were obtained after every set of
exposures on the stars.  After sky subtraction, the images were
flat fielded and corrected for hot pixels.  Aperture photometry was then
performed on the final images.

\subsection{Spectroscopy from the W. M. Keck Observatory}
Resolution $\sim$5000 spectra of stars B and C were taken with the LRIS
\citep{Oke} instrument on 1999 February 25 covering the spectral range of
6250$-$7550\AA.  A long slit of width 0.$''$7 was placed at a PA of
301.9 deg to obtain simultaneous spectra of both B and C, and two
integrations of 120~s each were made.  The amplifier bias was estimated
from the CCD over-read areas and subtracted from the spectra.  A
spectral flat field was made from halogen lamp exposures taken immediately
after the spectra and was divided into the spectra.  The seeing was 1.$''$1,
so B and C were not completely spatially resolved; their spectra were
deblended by fitting the spatial dimension at every spectral element
with two 1-D Gaussians plus a linear background.  The two observed
spectra of each star were then averaged.

High resolution spectra of all three stars were obtained with HIRES
\citep{Vogt94} on 1999 July 19.  The C1 decker was employed, which gives
a slit 0.$''$86 wide and 7$''$ long, yielding R=45,000 over 21 orders,
from 5420-7880\AA\ with some gaps.  The small separation of B and of C
required attention to scattered light.  Hence, the image rotator was
used to align the slit so that it was perpendicular to the vector
separating the stars.  Because of the good (0.$''$8) seeing, small slit
width, and care in position of the slit, we believe that the individual
spectra of B and C are uncontaminated by scattered light from the other
components.  A radial velocity standard, HR 6056 (-19.9 km s$^{-1}$), was
observed immediately following the spectroscopy of HD 141569 A/B/C.  The
spectra were extracted using the MAKEE package written by T. Barlow;
however, equivalent widths and radial velocities were measured using the
SPLOT package in IRAF.

\section{Results}

\subsection{Relative Motion}

\begin{figure}[tb]
\epsfig{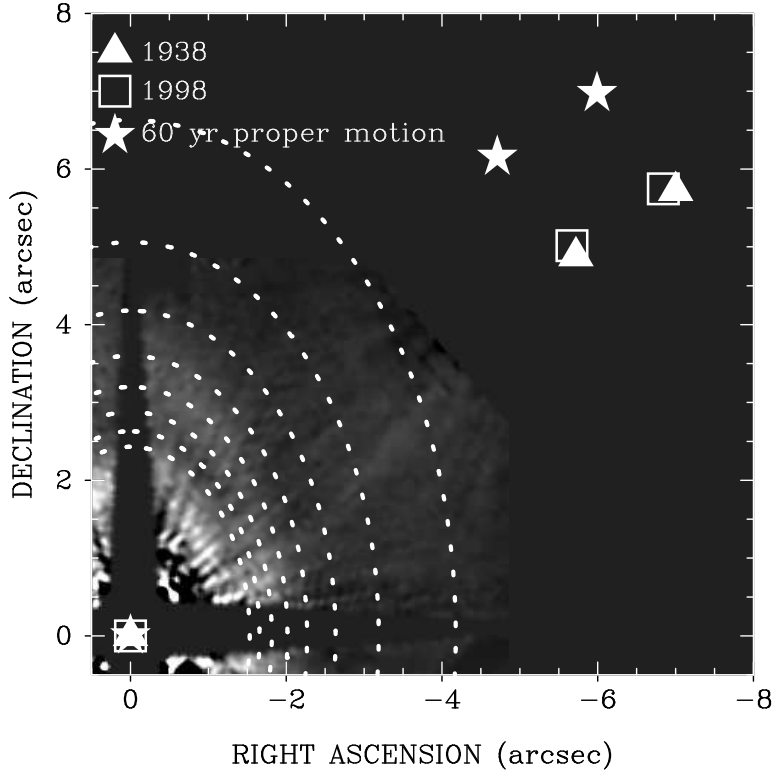}
\figcaption[f1.ps]{Positions of stars B and C relative to HD 141569A
shown for two epochs: 1938 (filled triangles) and 1998 (or 1999; open
squares).  Also shown are the predicted positions of B and C, if they
are stationary background stars, for 1998 given their 1938 locations and
the proper motion of A over the last 60 years (filled stars).  The sizes
of the points are larger than the uncertainties reported in Table 1 in
each case. The agreement between the 1938 and 1998 measurements shows
that A, B, and C are a co-moving group.  A NICMOS image of the disk
(Weinberger et al. 1999) is shown overplotted with dashed ellipses that
give the locations (from outside in) of the 2:1 -- 9:1 Lindblad
resonances between A--B/C and disk orbital periods assuming B and C lie
in the plane of the disk.  In projection, 1$''$ = 100 AU.
\label{fig_astromet}}
\end{figure}

A summary of measurements of relative separation and orientation of
HD~141569 A/B/C is given in Table \ref{table_astromet} and Figure
\ref{fig_astromet}.  The first point, from 1938 \citep{Rossiter55},
averages four measurements of B--C and three measurements of A--B made
in 1938 and 1943.  Although no uncertainties were given for the measured
separations and PAs, we used the scatter in the measurements to estimate
them.  The 1998 data are from our NICMOS acquisition images and the 1999
data from our Palomar images.  The uncertainties for the Palomar
positions were found from the standard deviation of many independent
measurements as described in \S\ref{photometry}. From 1938 to 1998, the
motions of the three stars are negligible to within the uncertainties,
and therefore they all have the same proper motion.


\begin{deluxetable}{llllll}
\tablenum{1}
\tablewidth{0pc}
\tablecaption{Summary of Astrometry \label{table_astromet}}
\tablehead{
\colhead{Date}
	&\colhead{Ref.} 
              &\colhead{A--B Sep}
		     &\colhead{A--B PA}
                     	 &\colhead{B--C Sep} &\colhead{B--C PA} \\
\colhead{}
	&\colhead{}
	      &\colhead{(arcsec)}
	             &\colhead{(deg)}
                     	      &\colhead{(arcsec)} &\colhead{(deg)}}
\startdata
1938	&Rossiter &7.5 $\pm$ 0.1 &310.5 $\pm$ 0.5 &1.5 $\pm$ 0.1 &302.7 $\pm$ 0.5\\
1998    &NICMOS   &7.57 $\pm$ 0.01 &311.5 $\pm$ 0.1  &1.38 $\pm$ 0.01 &301.9 $\pm$0.1\\
1999    &Palomar  &7.55 $\pm$ 0.02 &311.4 $\pm$ 0.1  &1.37 $\pm$ 0.01  &302.0 $\pm$ 0.3\\
\enddata
\end{deluxetable}

In 1995-1996, \citet{pirzkal} imaged HD 141569 A/B/C with shift-and-add
and measured the A--B separation as 6.$''$8 at a position angle of
312$^\circ$ and A--C as 8$''$.0 with PA=314$^\circ$.  No uncertainties
were provided in that paper, and these results are inconsistent with the
others.

\subsection{Spectral Types and Features}

\begin{figure}
\epsfig{file=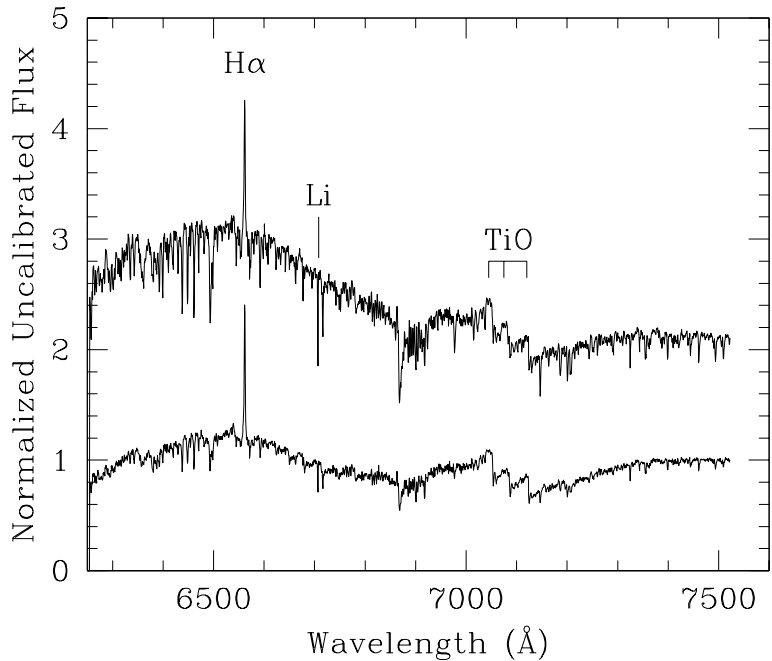,width=4in,clip=}
\figcaption[f2.ps]{Low resolution spectra of stars B (top) and C
(bottom).  Units of flux are arbitrary, since the spectra were not flux
calibrated, and global slopes are not meaningful.  Apparent are
H$\alpha$ in emission and Li 6708\AA\ in absorption as well as strong
7000\AA\ TiO bands. \label{fig_lowspectra}}
\end{figure}

\begin{figure}
\epsfig{file=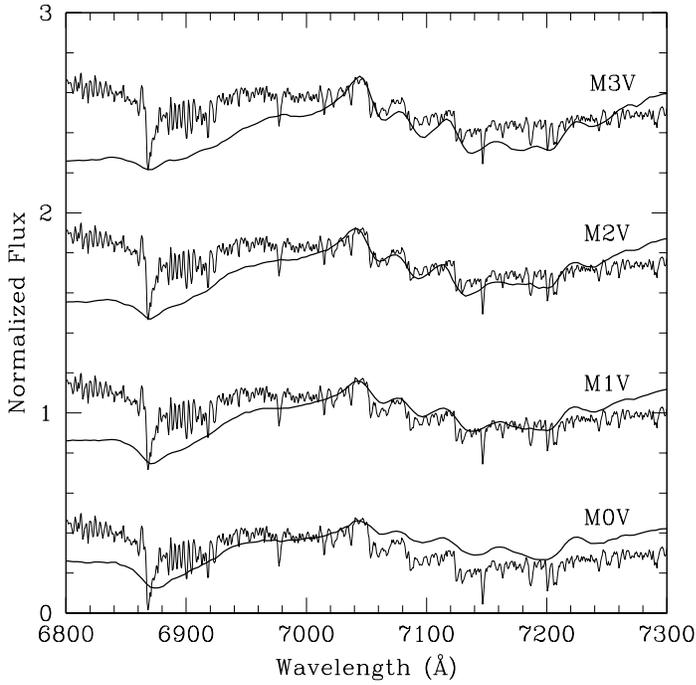,height=3.75in,clip=}
\figcaption[f3.ps]{Closeup of the spectrum of star B centered on the TiO
bands at $\sim$7000\AA\ compared to spectral standards of types M0~V --
M3~V from Kirkpatrick et al. (1991).  Again, global slopes are not
meaningful, since the spectrum was not divided by that of a spectral
calibrator.  The best matched standard is of type
M2~V.\label{fig_spectrumb}}
\end{figure}

\begin{figure}
\epsfig{file=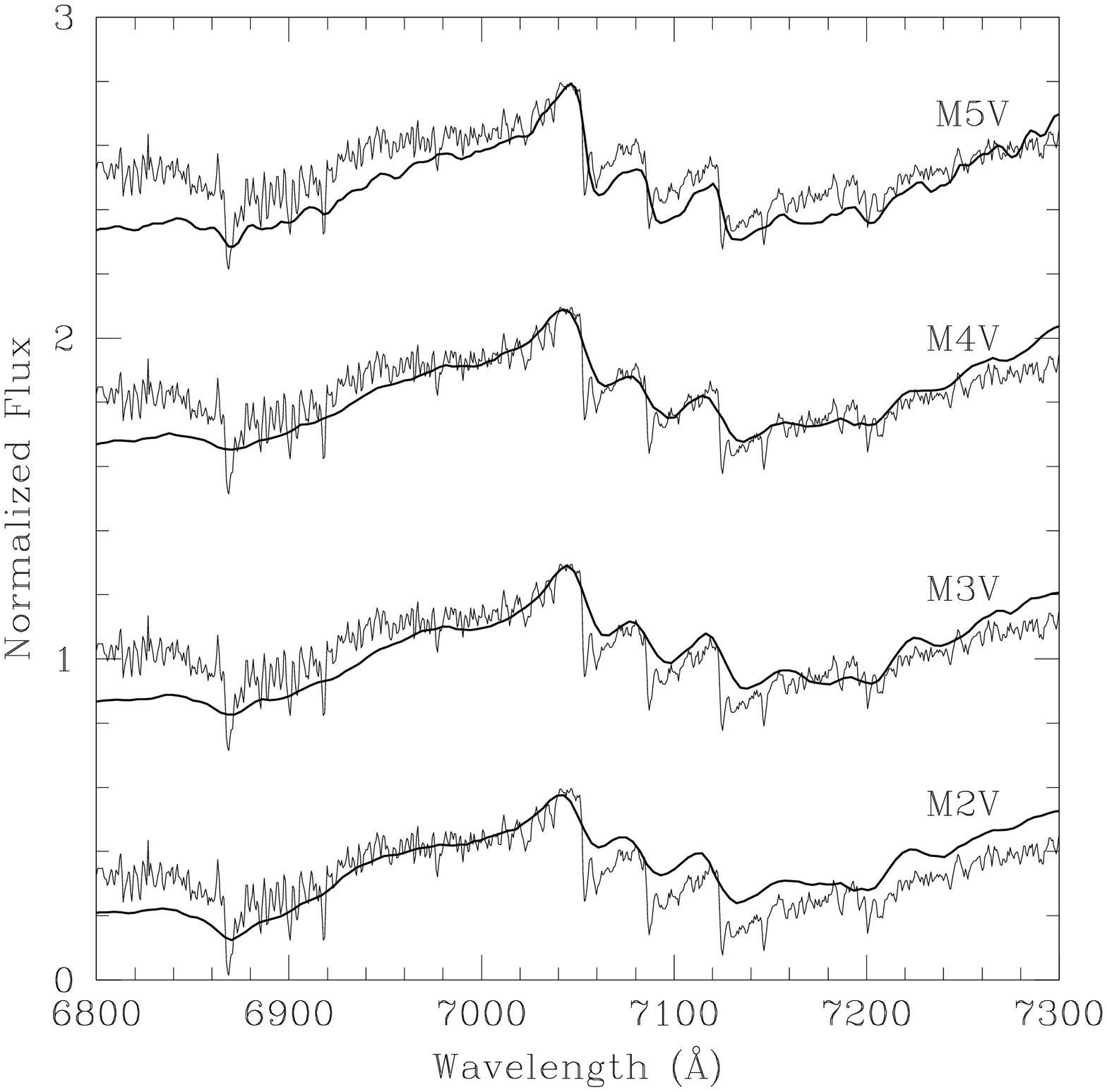,height=3.75in,clip=}
\figcaption[f4.ps]{Closeup of the spectrum of star C
centered on the TiO bands at $\sim$7000\AA\ compared to spectral
standards of types M2~V -- M5~V from Kirkpatrick et al. (1991).  The
best matched standard is of type M4~V. \label{fig_spectrumc}}
\end{figure}

The LRIS spectra of stars B and C are shown in Figure
\ref{fig_lowspectra}.  Notable features are H$\alpha$ in emission, Li~I
6708\AA\ in absorption and the TiO bands at $\sim$7000\AA\ which all
appear in both spectra.  The spectra have not been divided by a spectral
standard, so global slopes are not meaningful.  Figure
\ref{fig_spectrumb} presents a portion of the LRIS spectrum of B
compared to spectral standard stars in \citet{Kirkpatrick91}. Figure
\ref{fig_spectrumc} does the same for C.  We assign a spectral type of
M2~V to B and M4~V to C based on the depth of the TiO features.

\begin{deluxetable}{lccc}
\tablenum{2}
\tablewidth{0pc}
\tablecaption{Equivalent widths and heliocentric velocities\label{table_equivrv}}
\tablehead{
\colhead{Star}	&\colhead{H$\alpha$} &\colhead{Li I 6708\AA} &\colhead{V$_\odot$} \\
\colhead{}	&\colhead{EW (\AA)}   &\colhead{EW (\AA)} &\colhead{(km s$^{-1}$)} }
\startdata
A		&$-$5.51 $\pm$ 0.05	&\nodata             &$-$6 $\pm$5 \\
B		&$-$0.50 $\pm$ 0.04	&0.50 $\pm$ 0.04     &$-$1.3 $\pm$ 1.0 \\
C		&$-$1.70 $\pm$ 0.05	&0.50 $\pm$ 0.04     &$-$2.4 $\pm$ 1.1 \\
\enddata
\end{deluxetable}

\begin{figure}
\epsfig{file=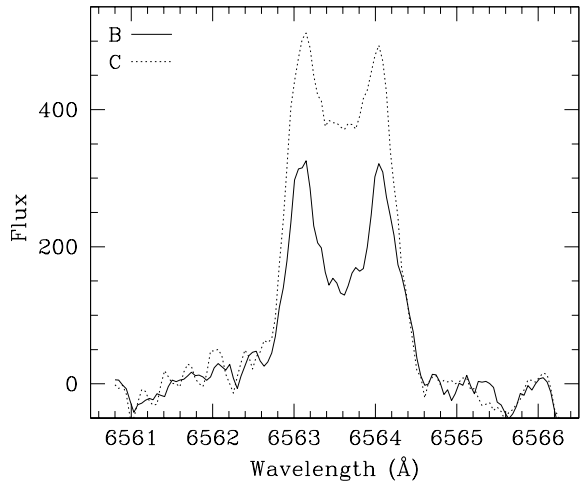,width=3in,clip=}
\figcaption[f5.ps]{H$\alpha$ line profiles of stars B and C after
continuum subtraction.  The double peaked structure, indicative of
chromospheric activity, is of the same width in both stars although the
line equivalent widths and depths of the central reversal
differ. \label{fig_halpha}}
\end{figure}

The equivalent widths of H$\alpha$ and Li~I are presented in Table
\ref{table_equivrv} and the H$\alpha$ line profiles for B and C are shown
in Figure \ref{fig_halpha}, both from the HIRES spectra.  Stars B and C
have double peaked lines with nearly the same shape and separations of
their peaks of 1\AA.  H$\alpha$ in star A is much broader,
double peaked with a peak to peak separation of 5.3\AA\ (242
km s$^{-1}$), and has a stronger blue than red peak; all of which are
consistent with previous measurements of H$\alpha$ by several authors
\citep{Andrillat,ZuckermanCP,Dunkin}.

\clearpage
\subsection{Radial Velocity}

From a cross-correlation of the high resolution B and C spectra over
each of the 21 spectral orders, B is moving away from us faster than C
by 0.9$\pm$0.4 km~s$^{-1}$.  The uncertainty is the standard deviation
in the mean of the 21 cross-correlations.  The radial velocity of B,
from cross correlation between its spectrum and that of the standard, is
--1.5 $\pm$ 0.6 km~s$^{-1}$.  We did not determine the radial velocity of
A because it had no lines suitable for comparison with the M0.5III
spectral-type radial velocity standard.  From the literature its
radial velocity is --6 $\pm$ 5 km s$^{-1}$ \citep{Frisch,Dunkin}.

\subsection{Photometry} \label{photometry}

In each photometric (i.e. 28 May) Palomar image, the magnitude of HD
141569A was measured in a 8$''$ diameter aperture.  In every short
non-photometric (i.e. 25 May) frame containing all three stars, A was
used as a point spread function (PSF) for fitting the locations of B and
C and flux ratios B/A and C/A, via minimization of the chi-square. All
40 images taken at J, and 20 images taken in each of the H and K bands
were fit independently.  Once the flux ratios were determined, the
magnitudes of B and C were calculated from the photometry of star A.
The results are summarized in Table \ref{table_phot}.  The uncertainties
reported are purely statistical and contain the statistical uncertainty
in the magnitude of A and the standard deviation of the independent
determinations of the flux ratios to B and C.  In the photometric L$_s$
image, the total flux from both B and C was measured in an 8$''$
diameter aperture centered between the stars and then star B was used as
a template PSF to fit the flux ratio of these two close stars.

\begin{deluxetable}{cccccc}
\tablenum{3}
\tablewidth{0pc}
\tablecaption{Photometry of HD 141569 System (Vega magnitudes)\label{table_phot}}
\tablehead{
\colhead{Star} &\colhead{J} &\colhead{H} &F171M  &\colhead{K$_s$} &\colhead{L$_s$}}
\startdata
A	&6.88	&6.84	&6.87	&6.83 &6.68 \\
B	&9.52	&8.82	&8.69	&8.64 &8.40 \\
C	&10.16  &9.44	&9.41	&9.25 &8.93 \\
\enddata
\tablecomments{Statistical uncertainties are 4\% at J, H, and K, 2--3\% at F171M, and 5\%
at L$_s$. No reddening corrections have been applied.}
\end{deluxetable}

\clearpage

\section{Discussion}

\subsection{Companionship and Associations}

The proper motion of component A has been measured by Hipparcos as
$-$16.86 $\pm$ 0.98 mas~yr$^{-1}$ in right ascension and $-$21.11 $\pm$
0.71 mas~yr$^{-1}$ in declination, so over the 60 years between
Rossiter's measurements and our own, it should have moved 1.$''$01 west
and 1.$''$27 south.  If B is a background star with negligible proper
motion, the PA of the A--B pair would have gone from 310.5\degr\ in 1938
to 323.5\degr\ in 1998 due to the proper motion of A ~(Figure 1).  A
change of this magnitude is ruled out by the measurements.  Furthermore,
as can be seen from Table \ref{table_equivrv}, all three stars have consistent
radial velocities.

If stars B and C were ordinary main sequence M-dwarfs, their photometric
distance would be $\sim$35 pc.  From the Hipparcos catalog, the average
proper motion of the 217 stars at this distance which are within 30 deg
of HD 141569A is 240 mas/yr, which is nearly nine times larger than that
of HD 141569A itself.  The relative separations of A--B and A--C would
have most likely changed by $\sim$15$''$ over 60 years if B and C were
foreground stars whereas in fact the separations have remained nearly
constant.

The measured changes in separation and position angle are consistent
with what would be expected from orbital motion.  Star B is at least 760
AU from A. If it is orbiting A, the orbital period of B is
$\gtrsim$13,500 yr and in 60 years it would move at most 1.6 deg in PA
(depending on the inclination of its orbit). Star C is at least 140 AU
from B.  If C is orbiting B, its period is $\gtrsim$2100~yr and in
60~yr, it would move at most 10 deg in PA.  The relative radial
velocities of B and C would be $<$2 km s$^{-1}$.  The actual changes in
PA (see Table \ref{table_astromet}) and the measured relative radial
velocities (see Table \ref{table_equivrv}) are well within these
constraints and imply that the stars could be orbiting one another.

Since A, B and C have common proper motions and common radial
velocities, they are with high probability physical companions.  Whether
the three stars are actually bound, however, cannot be determined.  As
noted in \citet{Weinberger99}, if the companion stars are in the plane
of the disk, the physical separation of A--B is 990 AU and that
of B--C is 190 AU.  Then, the ratio of the semi-major axes of
the wider to the closer pair would be only $\sim$5.2.  This is not
expected to be a stable triple system \citep{Eggleton95}, although such
young stars may not yet have had time to become unbound.

The presence of three young stars unassociated with a star-forming cloud
begs the question of how they formed.  HD 141569 is located in
projection near a complex of high latitude molecular cloud cores,
MBM~34-39 associated with the dark clouds L169, L183-4, and L134
(together called L134N).  The last of these has 100\micron\ emission
contours which actually encompass HD 141569 \citep{Sahu98}.  The excess
B--V color of HD 141569A, 0.095 mag, implies a reddening A$_v$=0.3 mag,
so the star cannot lie deep within the cloud. \citet{Sahu98} concluded,
based on the strength of interstellar absorption lines toward HD
141569A, that it lies behind part of L134N/MBM 37.  However, these
clouds have LSR radial velocities of 0.8--3.2 km s$^{-1}$
\citep{Magnani}, compared to $-$20 km s$^{-1}$ for HD 141569A, which
suggests that they are not co-moving with the star.  The clouds are also
quiescent and compact with internal temperatures of 3-12~K
\citep{snell81,clemens88}, which suggests that they are not undergoing
protostellar collapse or being heated by nearby stars.

The discovery of other young stars dispersed across the sky has prompted
speculation about fast cloud dispersal mechanisms and runaway
stars \citep{Feigelson}.  One way to address the question of star
formation in this region is to look for other nearby stars which may
have formed at the same epoch as HD 141569.  As a first attempt to
search for members of such an association, we have queried the Hipparcos
catalog using the approximate characteristics of the TW Hya association
\citep{Webb99}, a radius of 19 pc and a proper motion dispersion of 7
km s$^{-1}$, which at the distance to HD 141569 correspond to a circle of
radius 10\degr\ and a proper motion dispersion of $\sim$9 mas~yr$^{-1}$
in each direction.  In addition, we require the Hipparcos parallaxes to
agree with that of HD 141569A to within their uncertainties.  This
search produces 14 stars. In comparison, a search of 20 other fields at
the same absolute galactic latitude and a range of galactic longitudes
produces an average of 7 $\pm$ 5 stars using the same search parameters.
The data suggest an overdensity of stars near HD 141569.  We note that at
the distance of 100 pc, the Hipparcos catalog is highly incomplete for
stars fainter than V=9~mag (later than F).  

\begin{figure}
\epsfig{file=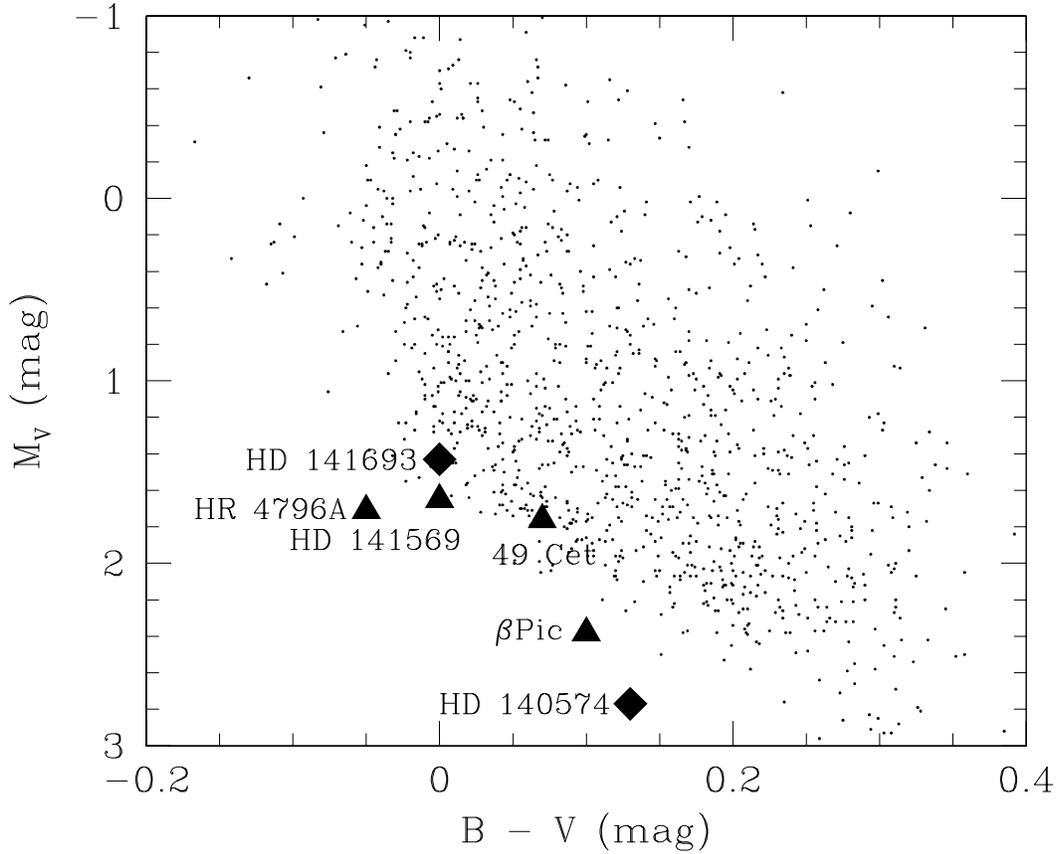,width=5.5in,clip=}

\figcaption[f6.ps]{All A type stars (small circles) from the Yale Bright
Star Catalog plotted as points on this Hertzprung-Russell diagram
reproduced from Jura et al. (1998).  Examples of young stars with
well-known infrared excesses are plotted with large triangles and fall
along the bottom of the distribution.  Two A type stars with the same
distance and proper motion as HD 141569, HD 141693 and HD 140574 (large
diamonds) also fall in this region of the diagram and thus also appear
to be young. \label{fig_HR}}
\end{figure}

Two of the 14 stars near HD 141569 are spectral type A and lie below or
along the ZAMS (Figure \ref{fig_HR}). This location in the HR diagram is
populated by young A type stars \citep{lowrance00} including well known
disk sources such as $\beta$ Pic and HR 4796A as well as HD 141569
itself.

We have computed the space motion of HD 141569A following
\citet{Johnson87} and find U,V,W = -3.0, -13.0, and -3.0 km s$^{-1}$.
This is within twice the velocity dispersion in the space motions of the
local star formation associations such as $\eta$ Chameleon, TW Hya, and
Tucanae \citep{Zuckerman00}.  So, the HD 141569 system may have formed
as part of a larger episode of star formation near the Sun.

\clearpage

\subsection{Age}

The pre-main sequence nature of stars B and C is confirmed by the
presence of Lithium in absorption. Since M-type stars should be fully
convective, Lithium in the spectra indicates that these stars are not
yet hot enough in their cores to burn it or have had insufficient time
to burn all of it.  The equivalent widths of 0.5\AA\ for B and C
is very close to the boundary of 0.54\AA\ set by \citet{Martin98}
for separating weak-line T Tauri stars (WTTS) from post-T Tauri stars.
This sets an upper limit on their age of $~$10~Myr \citep{Martin94}.  The
H$\alpha$ equivalent widths ($<$10\AA) also argue for classifying
these stars as WTTS as opposed to classical TTS.  The
strength of the double peaked H$\alpha$ emission is within the distribution of
chromospherically active main-sequence M dwarfs \citep{stauffer86}.  The
central reversal of the line indicates a high level of chromospheric
activity, which is generally associated with a variety of factors
including youth and rotation.  Finally, \citet{Lindroos85} classified
the combined B/C spectrum as ``peculiar'' because he detected hydrogen
and calcium emission lines, and pre-main-sequence stars often have such
emission lines.

The A/B/C system falls within the 90\% confidence error circle of a
ROSAT sky survey point-source detection.  The PSPC count rate of
0.093~s$^{-1}$ converts to an x-ray luminosity of
8.72~$\times$~10$^{29}$ erg~s$^{-1}$ assuming that it comes from stars
at 100~pc with T-Tauri-like x-ray spectra \citep{Neuhauser}.  It is
likely that the x-ray emission comes from the later-type stars, B and C
rather than from the primary star.  We estimate the bolometric
luminosities of stars B and C as log(L/L$_\odot$)=$-$0.62 and $-$0.95
respectively from their spectral types and J-band magnitudes using
bolometric corrections from \citet{Hartigan}.  This makes their total
x-ray to bolometric luminosity ratio log(L$_{\rm x}$/L$_{\rm
bol}$)=$-$3.2.  For pre-main sequence stars, the ratio of x-ray to
bolometric luminosity increases with stellar age, and the ratio for
stars B/C is typical for that of stars in clusters of age $<$20 Myr.
\citep{Kastner97}.

Finally, we estimate the age of stars B and C on the basis of their
location on theoretical pre-main sequence evolutionary tracks.  Here, a
major source of uncertainty is the not-well understood effective
temperature scale for M-dwarfs \citep{Allard97}.  Following the
spectral-type to temperature calibration of \citet{luhman98}, we can
assign an effective temperature for B (M2 V) and C (M4 V) of
3500$\pm$85~K and 3200$\pm$85~K, respectively, where the uncertainties
correspond to 1/2 spectral sub-class.  We note that the temperature
scale of \citet{Kirkpatrick93} gives a temperature 175~K hotter for
M4~V.

\begin{figure}
\epsfig{file=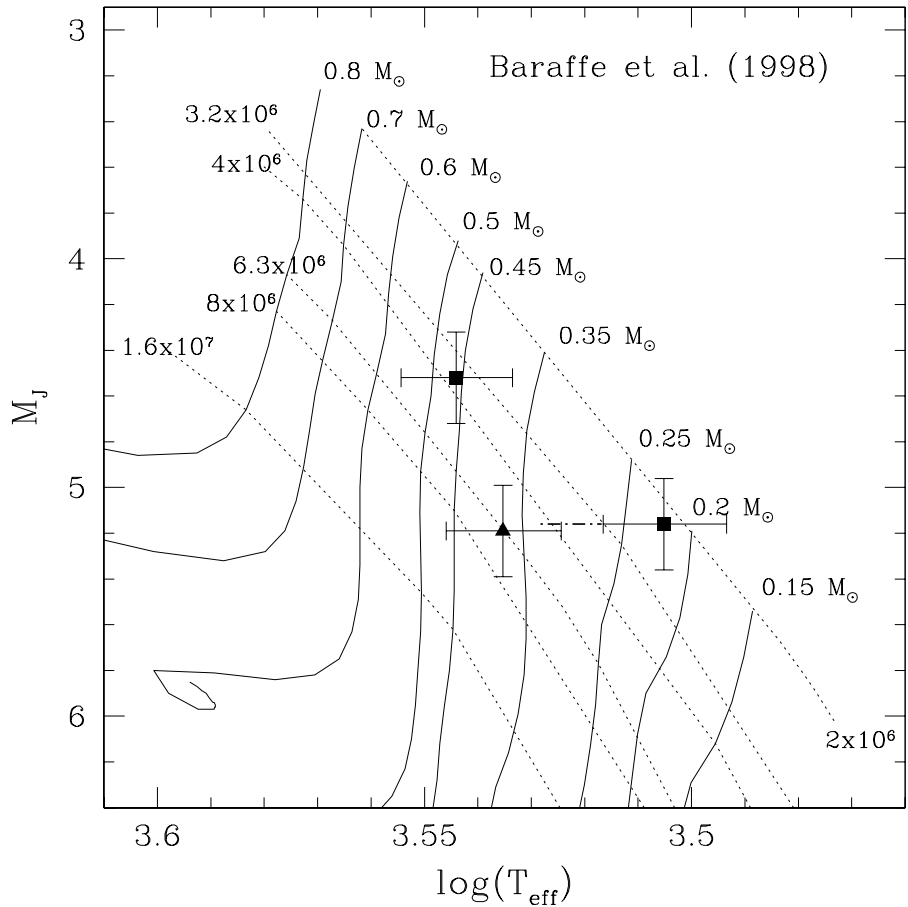,width=5in,clip=}

\figcaption[f7.ps]{Absolute J-band magnitudes and temperatures of stars
B (brighter) and C (fainter) plotted (filled squares) on pre-main
sequence tracks by Baraffe et al. (1998).  The same quantities are also
plotted for HR 4796B from Jura et al. 1993 (filled triangle).  The
effective temperatures and uncertainties thereon are from Luhman \&
Rieke (1998), but the dash-dot line on the effective temperature of C
shows the range of M4~V temperatures from other authors.  The age of HD
141569 B/C appears to be 2--8 Myr, about a factor of two younger than HR
4796B. \label{fig_tracks}}
\end{figure}

Shown in Figure \ref{fig_tracks} are tracks by \citet{Baraffe} with the
effective temperatures and absolute J-band magnitudes of stars B and C
plotted.  The plotted magnitudes assume that stars B and C are at the
same distance as A, 100~pc, and the uncertainties in the magnitudes
correspond to the uncertainty in the Hipparcos parallax to A.  Within
the uncertainties, stars B and C appear to be the same age of 2 to 8
Myr.  Star A is indistinguishable from the ZAMS and is not shown on the
figure.  The tracks indicate masses for stars B and C of 0.5 and 0.25
M$_\odot$, respectively.  Tracks by other authors give the same basic
result.  For comparison, Figure \ref{fig_tracks} also shows HR 4796B
with an effective temperature of a M2.5 star from \cite{luhman98}.  This
companion to HR 4796A, another star with a well studied circumstellar
disk, was assigned an age of 8$\pm$2 Myr by \citet{stauffer} using
tracks from \citet{dantona}.  The age given by the \citet{Baraffe}
tracks is consistent with that determination and shows that HD 141569
B/C are about a factor of two younger than HR 4796B.

Based on all of the above arguments, we estimate an age for the HD
141569 system of 5 $\pm$ 3 Myr.

\subsection{Dynamics}

The disk around HD 141569A has two features which suggest dynamical
sculpting \citep{Weinberger99}.  First, the density of scatterers is as
high at 360 AU as 200 AU from the star.  If the companions are out of
the plane of the disk, they could excite significant vertical velocities
in the disk dust.

Second, there is a dip in surface brightness, or a ``gap'' in the disk
at a radius of 250 AU with a width of 60 AU.  No point source is seen in
the gap to a limit of F110W=20.3 mag.  The gap is as circular as the
disk and must be cleared continually to remove particles drawn through
it by radiation pressure and Poynting-Robertson drag.  If the companion
stars orbit each other and the primary and are in the plane of the disk,
their center of mass is 1053 AU from the primary.  The 2--9:1 Lindblad
resonances between the orbital period of the companion center of mass
and orbiting particles in the disk are shown with the dashed ellipses on
Figure \ref{fig_astromet}. There is no obvious agreement between the
resonances and the structure observed in the disk, even if the
companions are assumed to be somewhat out of the plane and so the
resonances shifted.  The resonances closest to the gap lie at 243 AU
(9:1) and 263 AU (8:1) from the star.  It is not clear, however, why
only these high order resonances and not any of the others would mold
the disk.  If the companions were in very eccentric orbits and currently
near apastron, the locations of the resonances might change dramatically
with time. However, the circularity of the gap implies a stable
dynamical influence over long time periods.

\section{Conclusions}

On the basis of common proper motion, common radial velocity and the low
probability of a chance superposition of three young stars away from any
known star-forming cloud, we conclude that HD 141569 A/B/C form a
physical association which may or may not be bound.  The age of the
stars as determined from spectroscopic features, x-ray emission, and
placement on pre-main sequence tracks is 5  Myr.  At least two
other stars, HD 141693 and HD 140574, may be members of this common
proper motion group.

HD 141569 has a now well studied disk with grains present from 25 -- 500
AU of the star.  The inferred size of the grains, $<$5 $\mu$m
\citep{Fisher00}, means that radiation pressure blows them away in at
least an order of magnitude less time than our derived stellar age.
Thus, they must be continuously regenerated, probably through collisions
of larger bodies.  The timescales for the formation of planetesimal
cores is 10$^4$ - 10$^5$ yr \citep{Wetherill} which is easily consistent
with the age of the system.  From measurements of CO around HD 141569,
the remnant mass of H$_2$ is estimated as 20 to 460 M$_{\rm Earth}$
\citep{Zuckerman95}, so any gas giants present must have formed very
quickly.

The morphology of the disk, including a gap at 250 AU, indicates the
presence of dynamical sculpting.  Resonant interactions between the
companions and the disk do not appear to account for the structure, and
it would be hard given current models of planet formation to generate
Jupiter sized bodies in a disk of such young age at distances so far
from the central star \citep{Boss98}.  The cause of the structure in the
disk thus remains unknown.

\acknowledgements

We thank G. Neugebauer for donating 200-inch Telescope time to this
project. We thank B. Schaefer and R. Quick for making the LRIS
observations at the W. M. Keck Observatory, which is operated as a
scientific partnership between the California Institute of Technology,
the University of California, and the National Aeronautics and Space
Administration and was made possible by the generous financial support
of the W. M. Keck Foundation.  This work is supported in part by NASA
grant NAG 5-3042, and based on observations with the NASA/ESA Hubble
Space Telescope, obtained at the Space Telescope Science Institute,
which is operated by the Association of Universities for Research in
Astronomy, Inc. under NASA contract NAS5-26555. We also thank M. Jura
for many helpful discussions.

\end{document}